\documentclass[prb,twocolumn,amssymb,amsmath,floatfix,aps,10pt]{revtex4-1}

\usepackage{graphicx}

\usepackage{bm}

\usepackage{lipsum}

\usepackage{threeparttable}

\usepackage[english]{babel}
\usepackage[utf8x]{inputenc}
\usepackage[T1]{fontenc}

\usepackage{graphics}
\usepackage{graphicx}
\usepackage{subfigure}
\usepackage{epsfig}
\usepackage{etoolbox}
\usepackage{setspace}
\usepackage{nonfloat}
\usepackage{lastpage} % Get access to the number of pages
\usepackage{appendix}

\usepackage{parskip} % remove indentation from new paragraphs
\usepackage{amsmath}
\usepackage[colorinlistoftodos]{todonotes}
\usepackage[colorlinks=true, allcolors=blue]{hyperref}
\usepackage{hhline}

\addto\captionsenglish{}
\addto\captionsenglish{}

\usepackage{soul}

\begin{document}
\title{Database-driven high-throughput study for hybrid perovskite coating materials}
\author{Azimatu Seidu}
\affiliation{Department of Applied Physics, Aalto University, P.O.Box 11100, FI-00076 AALTO, Finland}
%\email{azimatu.seidu@aalto.fi}
\author{Lauri Himanen}
\affiliation{Department of Applied Physics, Aalto University, P.O.Box 11100, FI-00076 AALTO, Finland}
\author{Jingrui Li}
%\email{jingrui.li@aalto.fi}
\affiliation{Department of Applied Physics, Aalto University, P.O.Box 11100, FI-00076 AALTO, Finland}
\author{Patrick Rinke}
\affiliation{Department of Applied Physics, Aalto University, P.O.Box 11100, FI-00076 AALTO, Finland}

\begin{abstract}
We developed a high-throughput screening scheme to acquire candidate coating materials for hybrid perovskites. From more than 1.8 million entries of an inorganic compound database, we collected 93 binary and ternary materials with promising properties for protectively coating  halide-perovskite photoabsorbers in perovskite solar cells. These candidates fulfill a series of criteria, including wide band gaps, abundant and non-toxic elements, water-insoluble, and small lattice mismatch with surface models of halide perovskites.
\end{abstract}
\maketitle       

Perovskite solar cells (PSCs) \cite{snaith,green,Saliba18} have recently reached a power-conversion efficiency (PCE) of $>\!\!23\%$ only six years after the invention of the state-of-the-art PSC architecture in 2012 (PCE$\sim$10\%) \cite{kim,lee}. This has revived the hope for direct conversion of sustainable, affordable and environmentally friendly solar energy into electricity. The photoabsorbers in PSCs are hybrid (organic-inorganic) perovskites (denoted $\text{ABX}_3^{}$ hereafter) especially methylammonium (MA) lead iodide ($\text{CH}_3^{}\text{NH}_3^{}\text{PbI}_3^{}\!\equiv\!\text{MAPbI}_3^{}$). The salient properties of these materials in optoelectronic applications are optimal band gaps, excellent absorption in the visible range of the solar spectrum, good transport properties for both electrons and holes, flexibility of composition engineering, as well as low costs in both raw materials and fabrication \cite{snaith,green,stranks,Xing13,eperon,troughton}.

Despite the excellent PSC-performance in the laboratory, stability problems limit the development and commercialization of this promising materials class. Hybrid perovskites degrade quickly in heat, oxygen and moisture \cite{niu,Niu15,HuangJ17,Mesquita18,Ciccioli18}. With  increasing exposure to any of these destabilizing factors, the structure of the hybrid perovskite degrades and the PCE reduces concomitantly  after several days or even hours \cite{KimGH17,LiF18}. Among the solutions that have been proposed to  solve this stability and longevity problem are protective coating \cite{Matteocci16,Cheacharoen18a,Cheacharoen18b}, the use of two-dimensional perovskites \cite{Quan16,Dou17,Ran18a,WangZ18}, and doping with small ions \cite{noh,Yi16,ZhouY16,Tan17,Ciccioli18}. Protective coating is particularly promising, as it can passivate the surface dangling bonds of the perovskite photoabsorber and insulate the perovskite from heat and small molecules from the environment. A good coating should have the following properties: (i) a wide band gap (>3 eV), (ii) little impact on the structure of the coated perovskite, (iii) good transport properties, and (iv) high stability in heat, air and water. It would be particularly attractive, if the coating material could also be used as a semiconducting interlayer, a key component in the modern perovskite-based device architectures. In this context, we are especially interested in cheap and efficient hole-transporting coatings, as \textit{Spiro}-OMeTAD, the most common hole-transporting material (HTM) in PSCs since the birth of this technology \cite{kim,lee}, is expensive, has low charge-carrier mobilities and a negative impact on PSC stability \cite{Saliba16b}.

We here present a database-driven high-throughput study that explores a wide range of possible candidates to find inorganic materials that have the potential to protectively coat perovskites in PSCs. We take the inorganic materials from the ``Automatic Flow for Materials Discovery'' (\textsc{aflow}) database \cite{stefano}. \textsc{aflow} contains nearly 2 million material entries that were computed with density-functional-theory (DFT) using the Perdew-Burke-Ernzerhof (PBE) exchange-correlation functional \cite{perdew}.

\begin{figure}[!ht]
\vspace{-0.5em}
\includegraphics[clip=true,trim=0.7in 6.1in 0.7in 1.1in,width=0.48\textwidth]{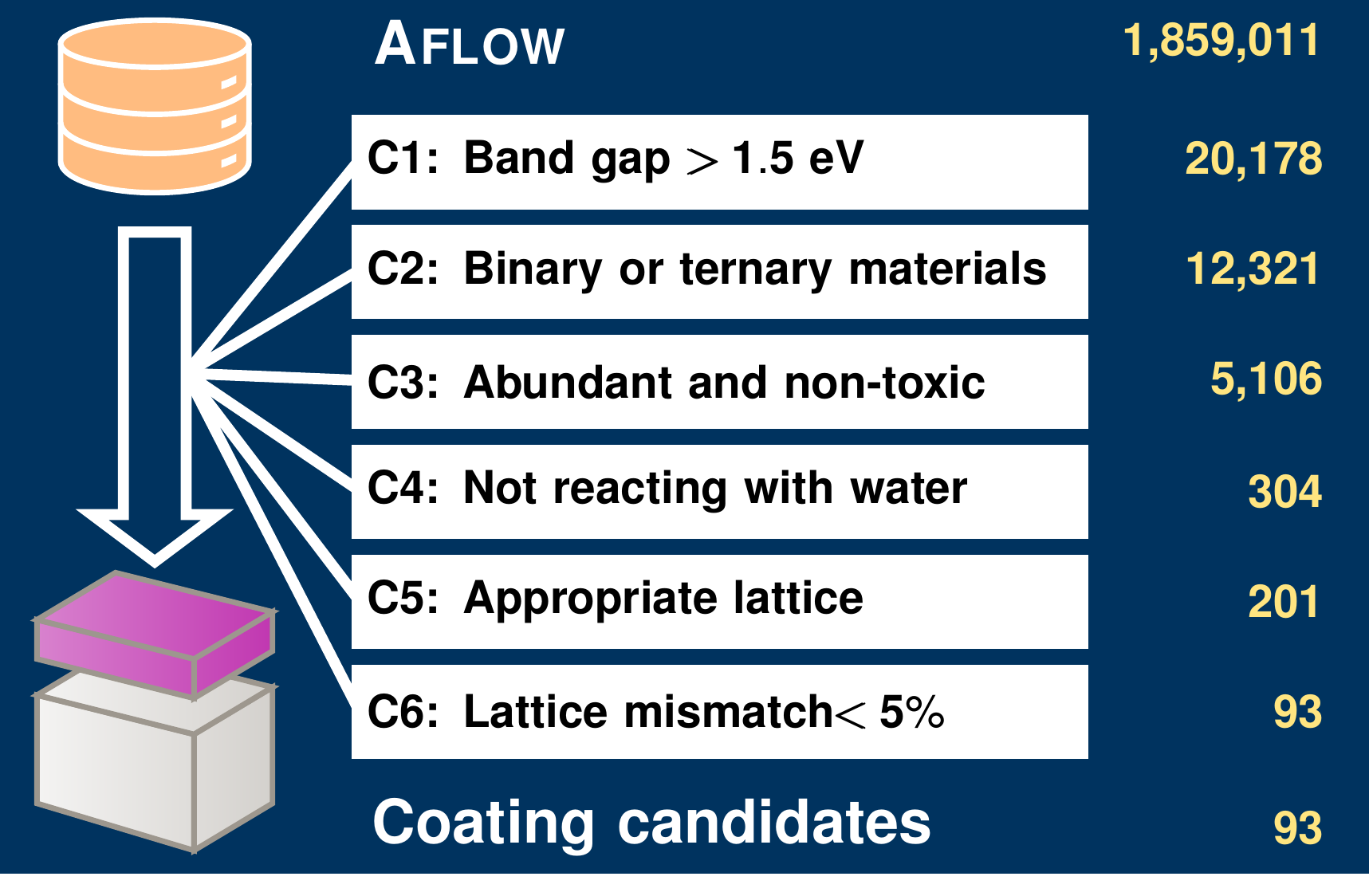}
\vspace{-2em}
\caption{High-throughput screening scheme to extract possible coating materials from \textsc{aflow}. The six filtering criteria are listed in the middle and the corresponding number of remaining compounds is given by the numbers in yellow.}  \label{fig:query} \end{figure}

In the following, we will describe our filtering scheme with which we reduced the large number of database entries to only those material candidates with promising coating properties. The workflow is illustrated in Fig.~\ref{fig:query}. Since PBE generally underestimates band gaps by $\sim\!\!50\%$ \cite{sham83}, we set our first criterion (C1) to screen materials with ``PBE band gap $>\!\!1.5~{\text{eV}}$''. Considering the technical difficulties of coating with quaternary or even more complicated compounds \cite{Ryou09,Yeh14}, we limited our target materials to binary and ternary compounds in this work (C2). In C3 we excluded all compounds that contain toxic or rare elements, and in C4 we discarded the compounds that are unstable in contact with water. Details of how we implemented C3 and C4 are available in the Supplementary Material (SM). In C5, we selected candidates with appropriate lattices, meaning candidates with at least two perpendicular lattice vectors in the conventional cell. In the final step (C6), we calculated the lattice mismatch between selected perovskite substrates and the coating materials that survived from C5. This last step produced some coating materials with several phases. In such cases, we prioritized the phase with the least lattice mismatch to $\text{MAPbI}_3^{}$. The other crystal phases are presented in the SM.

\begin{figure}[!ht]
\includegraphics[page=1,clip=true,trim=0.8in 7.3in 0.9in 1.0in,width=0.45\textwidth]{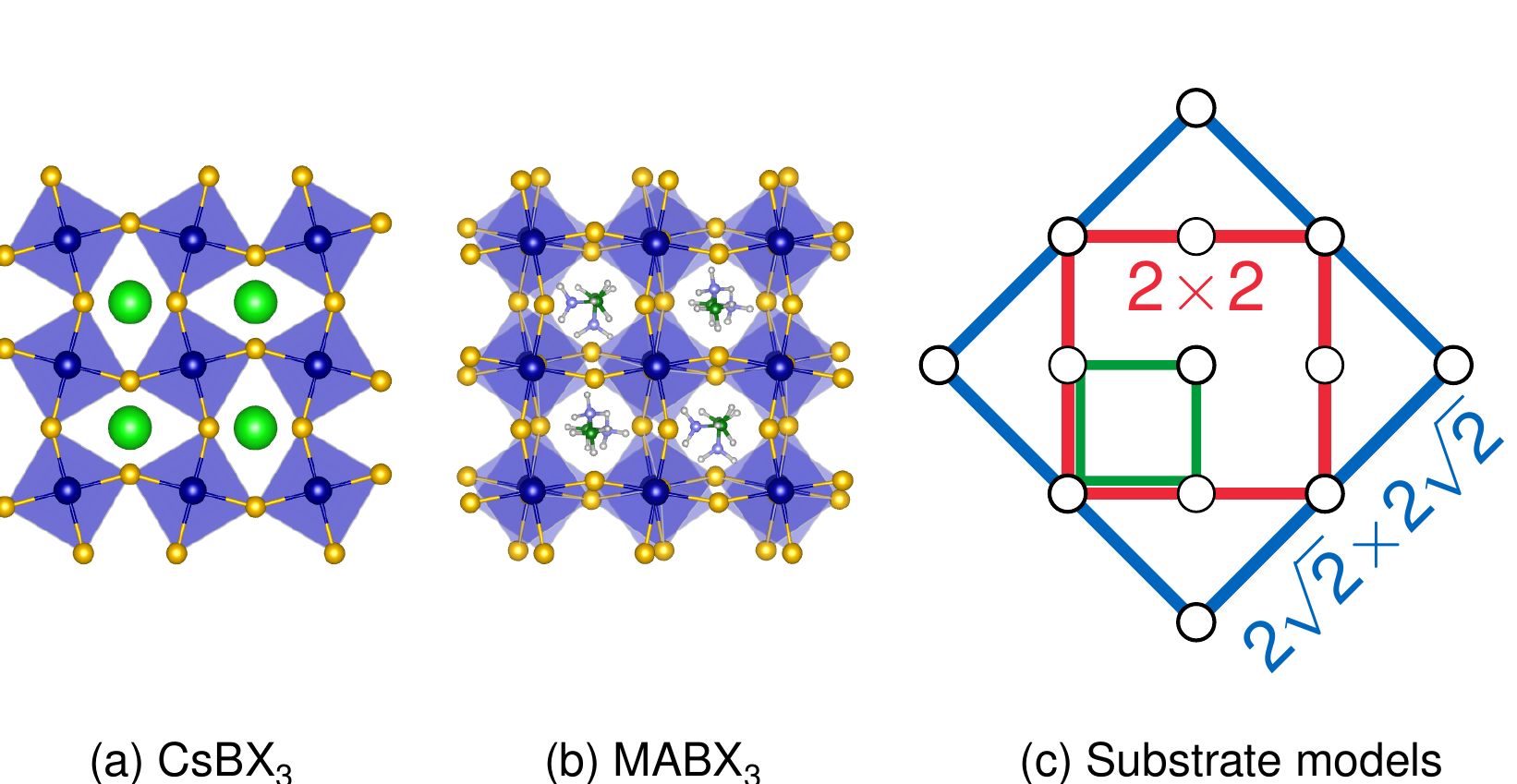}
\caption{$(001)$ plane of tetragonal  $\text{CsBX}_3^{}$ (a) and  $\text{MABX}_3^{}$ (b). The red square in (c) depicts the $2\times2$ and the blue the $2\,\sqrt{2}\times2\,\sqrt{2}$ unit cell in the $(001)$ plane of of $\text{ABX}_3^{}$. The green square denotes the square primitive cell. Empty circles indicate the lattice points (e.g., B-sites) at the $(001)$ plane.} \label{fig:model}
\end{figure}

As substrates, we chose 12 $\text{ABX}_3^{}$ perovskites (A = Cs/MA, B = Sn/Pb, and X = Cl/Br/I) that are commonly used in halide-perovskite-based devices. We optimized the structure of the tetragonal P4/mbm phase of $\text{CsBX}_3^{}$ and the tetragonal (quasi) I4/mcm phase of $\text{MABX}_3^{}$ using PBE \cite{perdew} (to stay consistent with \textsc{aflow} data \cite{taylor}) and the analytic stress tensor \cite{knuth} implemented in the all-electron numeric-atom-centered orbital code \textsc{fhi-aims} \cite{blum09,havu,Levchenko/etal:2015}. Details of the DFT calculations are given in the SM. Upon a test calculation, we selected the $(001)$ crystal planes of the perovskites (Figs.~\ref{fig:model}a and b) since they are the most stable surface of these materials. We determined the lattice mismatch based on the lattice constants alone and did not carry out any interface calculations with DFT. Figure~\ref{fig:model}c shows the two ``virtual surface models'' considered in this work. We did not consider larger surface models, since they would make further computational modeling intractable.

From the PBE-optimized lattice constants, we calculated the lattice mismatch at each coating-perovskite interface. To avoid large strain, we required that the coatings should have rectangular lattice planes with small miller indices, e.g., the $(100)$ plane of the cubic lattice or the $(11\bar{2}0)$ plane of the hexagonal lattice. More details of this selection is given in the SM. If the lattice constant of the coating and the perovskite are $a_{\text{c}}^{}$ and $a_{\text{p}}^{}$ along one direction, then the lattice mismatch is,

\begin{align}
\gamma \triangleq \frac{ma_{\text{c}} - a_{\text{p}}}{ma_{\text{c}}} \times 100\%, &\quad m\in\mathbb{N}. \label{eqn:coat}
\end{align}

$m$ is the integer that minimizes $\vert\gamma\vert$. We set the criterion $\gamma\in(-5,+5)\%$ as shown in Fig.~\ref{fig:query}.
 
With the high-throughput screening scheme in Fig.~\ref{fig:query},  we extracted 93 inorganic semiconductor coating candidates (39 binaries and 54 ternaries) from \textsc{Aflow}. In addition, there are $\sim$1000 suitable ternary compounds, for which we could not find any data on their solubility in water. These remaining compounds will be investigated further in the future.

\begin{figure*}[!ht]
\includegraphics[width=17.5cm]{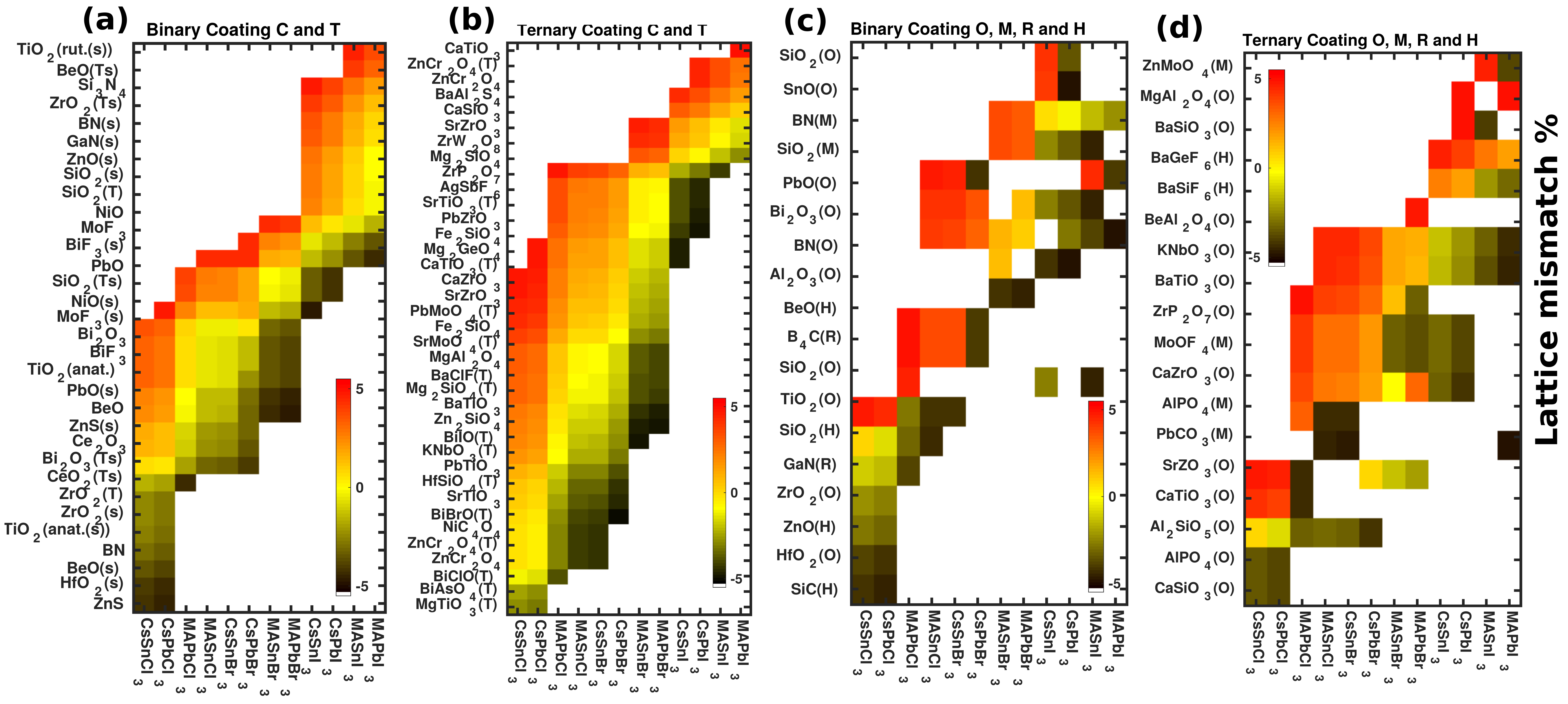}
\caption{Calculated lattice mismatch ($\gamma$) between the considered perovskites (horizontal axes) and suitable coating materials (vertical axes). \textsc{c},  \textsc{t}, \textsc{o}, \textsc{m}, \textsc{h} and \textsc{r} are short for cubic, tetragonal, orthorhombic, monoclinic, hexagonal and rhombohedral crystal structures, respectively. s denotes  $2\sqrt{2}\times2\sqrt{2}$ perovskite substrates, all others are $2\times2$. Panel shows (a) binary \textsc{c} (unlabeled) and \textsc{t} coatings, (b) ternary \textsc{c} and \textsc{t} coatings, (c) and (d) ``non-square'' (i.e., \textsc{o}, \textsc{m}, \textsc{h} and \textsc{r}) for binary and ternary coatings.}
\label{fig:plot}
\end{figure*}

Figure~\ref{fig:plot} shows the calculated lattice mismatch between the candidates and the 12 $\text{ABX}_3^{}$ perovskite substrates. Panels~\ref{fig:plot}a and b reveal that several materials with cubic or tetragonal lattices can  be used to coat most of the investigated perovskites: $\text{ZnS}$, $\text{BN}$, some fluorides ($\text{BiF}_3^{}$, $\text{MoF}_3^{}$ and $\text{AgSbF}_6^{}$), some binary oxides ($\text{Bi}_2^{}\text{O}_3^{}$ in both cubic and tetragonal phases, $\text{Ce}_2^{}\text{O}_3^{}$, $\text{BeO}$, $\text{PbO}$, $\text{TiO}_2^{}$-anatase, $\text{NiO}$ and tetragonal $\text{SiO}_2^{}$) and a large range of ternary oxides. In contrast, Figures.~\ref{fig:plot}c and d show that most of the materials that are in neither the cubic nor the tetragonal phase can only cover a small range of perovskite substrates. This is because the $\vert\gamma\vert<5\%$ criterion must be satisfied by two lattice constants, which makes the coating less ``versatile'' in these phases.

From Figs.~\ref{fig:plot}a and b, one can immediately deduce that the lattice mismatch increases from $-5$ to $5$ \% as the lattice constant of the substrates increases. The yellow spots show the most promising candidates with mismatch $<1\%$. Only a few coating candidates with ``non-square'' planes survived our screening criteria. This is because in such materials, at least two lattice constants must  have lattice mismatch within $-5$ and $5\%$. For instance, the $\gamma$ values for the interface between the hexagonal phase of $\text{Bi}_2^{}\text{O}_3^{}$ at $\text{MAPbBr}_3^{}$ interface are $-6.5\%$ and $0.73\%$ along the $a$- and $c$-axis, respectively. Thus $\text{Bi}_2^{}\text{O}_3^{}$ would not be a suitable candidate to coat $\text{MAPbBr}_3^{}$. 

As a first consistency check, we compared  the material candidates in Figs.~\ref{fig:plot}a and b to materials that have already been used as transport or mesoporous scaffold layers in PSCs. We found that our search is consistent with common materials such as: $\text{NiO}$ as  HTM in PSCs \cite{lai18}, as well as $\text{ZnO}$ \cite{lai18} and $\text{TiO}_2^{}$  \cite{chen14} as electron-transporting materials (ETMs). Similarly, our candidate materials included $\text{ZrO}_2^{}$ \cite{escobar17} and $\text{Al}_2^{}\text{O}_3^{}$ \cite{Si16} which are used as mesoporous scaffolds in PSCs.  

Aside from the commonly known metal oxides used in PSCs, we discovered some surprising binary candidates ($\text{MoF}_3^{}$, $\text{GaN}$, $\text{BiF}_3^{}$, $\text{Si}_3^{}\text{N}_4^{}$ and $\text{BN}$) that have properties suitable to coat the photovoltaic-active halide perovskites (Fig.~\ref{fig:plot}a). Similarly, for ternaries we found $\text{BaAl}_2^{}\text{S}_4^{}$, $\text{AgSbF}_6^{}$, $\text{BaSiF}_6^{}$ and $\text{BaGeF}_6^{}$. These materials came as surprise since they are usually not considered in PSCs due to their high melting temperatures. However, with new coating techniques such as radio-frequency sputtering \cite{Clemente18}, pulsed laser deposition \cite{Liang16}, vapor-deposition \cite{Jorge17} and modified hybrid methods such as  spin-coating/vapor-deposition \cite{Dong15}, these materials become contenders as effective coating materials for future PSC devices. 

Of particular interest are the potential coating materials for $\text{MAPbI}_3^{}$, the most common photoabsorber in PSCs. Interestingly, our screening procedure reveals that $\text{Al}_2^{}\text{O}_3^{}$ (Fig.~\ref{fig:plot}c), which is the most common mesoporous material in today's PSC architectures \cite{lee}, does not have the minimum lattice mismatch for coating  $\text{MAPbI}_3^{}$. $\text{ZnO}$, $\text{NiO}$, $\text{CaSiO}_3^{}$, $\text{SiO}_2^{}$, $\text{SrZrO}_3^{}$, $\text{BaAl}_2^{}\text{S}_4^{}$, $\text{GaN}$, $\text{MoF}_3^{}$, $\text{BN}$, $\text{Si}_3^{}\text{N}_4^{}$ and $\text{ZrO}_2^{}$ lead to better lattice match. The actual strain values for $\text{MAPbI}_3^{}$ can be found in the far right column of each panel in Fig.~\ref{fig:plot} 

Next we briefly address the charge carrier properties of the potential candidates. Table~\ref{tab:MAPI} lists the PBE band gaps of the found candidate coatings for $\text{MAPbI}_3^{}$ provided by \textsc{Aflow} \cite{stefano}, together with the dominant charge carrier type (n- or p-type). Here, we observe that intrinsic p-type semiconductors such as NiO and PbO, will not only protect PSCs against ambient conditions, but could also serve as efficient HTMs to replace the inefficient \textit{Spiro}-OMeTAD.

We also found insulators such as $\text{ZrO}_2^{}$, $\text{Si}_3^{}\text{N}_4^{}$, and $\text{BeO}$  (Table~\ref{tab:MAPI}). Due to the large band gap of these materials and their insolubility in water, they can be used as efficient mesoporous scaffolds to  passivate  PSCs against degradation. Additionally, BN could be used as a p-- or n--type semiconductor with different doping mechanisms (Table~\ref{tab:MAPI}). It was recently reported that BiF$_3$ has a high-lying valence band \cite{feng, Poole76} thus potentially being a good HTM.  Also HfO$_2$ could be engineered into a p-type material by controlling the oxygen vacancy content \cite{Hildebrandt11}.

\begin{table*}[!ht]
\caption{Selected candidate coating materials for $\text{MAPbI}_3^{}$. Listed are their band gaps (in eV, data from \textsc{Aflow}\cite{stefano}), space groups, and dominant charge carriers (n- or p-type conductivity) with corresponding references. Intrinsic n- and p-type materials, and n- and p-type dopable materials, are labeled by N, P, n and p, respectively.}\label{tab:MAPI}
\begin{tabular}{@{\hspace{0.2em}}
ccccc@{\hspace{0.2em}}p{1em}@{\hspace{0.2em}}
ccccc@{\hspace{0.2em}}} \hline\hline
Coating & Space group & Gap  & Cond. & Refs. & ~ &
Coating & Space group & Gap  & Cond. & Refs. \\ \hline
CaSiO$_3$  & P4\=3m  & $3.67$ &  & & &
PbO        & P4/nmm     & $1.66$  & P & \onlinecite{Droessler14} \\
BaAl$_2$S$_4$  & Pa\=3      & $3.65$ & & & &
ZnO        & F\=43m     & $1.69$  & N & \onlinecite{Stevanovic14} 
\\
SiO$_2$    & Fm\=3m     & $1.73$  & & & & 
CaZrO$_3$  & Pm\=3m      & $3.21$ & & 
\\
MoF$_3$    & Pm\=3m     & $1.74$ & & & &
GaN        & F\=43m     & $1.75$ & &  
\\
NiO        & Fm\=3m     & $1.79$  & P & \onlinecite{Stevanovic14} &&
ZrO$_2$    & P4$_2$/nmc & $3.86$ & & 
\\
BiF$_3$    & Fm\=3m      & $3.95$ & & \onlinecite{Poole76} & &
HfO$_2$    & P4$_2$/nmc & $4.01$ & p & \onlinecite{Hildebrandt11}
\\
BaTiO$_3$  & Pm\=3m     & $2.13$ & & & & 
BN         & F\=43m     & $4.46$ & n,p &
\\
BN         & P6$_3$mc   & $5.21$ & n,p & & &
PbZrO$_3$  & Pm\=3m      & $2.28$ & & 
\\
HfSiO$_4$  & I4$_1$/amd & $5.31$ & & & & 
CaTiO$_3$  & Pm\=3m      & $2.36$ & & 
\\
BeO        & Fm\=3m     & $8.18$ & & & &
Si$_3$N$_4$ & Fm\=3m & $3.33$ & &
\\ \hline\hline
\end{tabular}
\end{table*}

Lastly, we briefly comment on realistic coating interfaces. The actual phase of the coating material and the structure of the interface depend on many factors such as the perovskite surface structure and properties, the deposition method, the deposition conditions, as well as the coating thickness. These factors are not included in our database study. An atomistic description of coating-perovskite interfaces requires further computational (e.g., DFT) and experimental work. Results from such future work, such as the stability of the coating materials, could then be incorporated as additional criteria in our screening procedure. 

In summary, we have developed a systematic and efficient screening scheme for perovskite coating materials. Our scheme reduces the $\sim$1.8 million materials entries in \textsc{Aflow} to 93 possible coating candidates for a series of perovskite photoabsorbers in PSCs.  We have identified inexpensive HTMs (NiO and PbO) that can replace the inefficient and expensive \textit{Spiro}-OMeTAD, as well as several efficient ETMs (e.g., ZnO) for PSCs. Our results  feature new materials beyond metal oxides that will not only enhance the stability of PSCs  but also serve as a starting point in the search of novel device materials for emergent PSC technologies.

We gratefully thank M. Todorovi\'c and G.-X. Zhang for insightful discussions.  We acknowledge the computing resources by the CSC-IT Center for Science and the Aalto Science-IT project. An award of computer time was provided by the Innovative and Novel Computational Impact on Theory and Experiment (INCITE) program. This research used resources of the Argonne Leadership Computing Facility, which is a DOE Office of Science User Facility supported under Contract DE-AC02-06CH11357. This project has received funding from the European Union's Horizon 2020 research and innovation programme under grant agreement No 676580 with The Novel Materials Discovery (NOMAD) Laboratory, European Center of Excellence, the V\"ais\"al\"a Foundation, as well as the Academy of Finland through its Centres of Excellence Programme (2015-2017) under project number 284621 and its Key Project Funding scheme under project number 305632. 

\bibliographystyle{apsrev}
\bibliography{sample}

\end{document}

\title{{\normalsize Supplementary Material for} \\ Database-driven high-throughput study for hybrid perovskite coating materials}
\author{Azimatu Seidu}
\affiliation{Department of Applied Physics, Aalto University, P.O.Box 11100, FI-00076 AALTO, Finland}
%\email{azimatu.seidu@aalto.fi}
\author{Lauri Himanen}
\affiliation{Department of Applied Physics, Aalto University, P.O.Box 11100, FI-00076 AALTO, Finland}
\author{Jingrui Li}
%\email{jingrui.li@aalto.fi}
\affiliation{Department of Applied Physics, Aalto University, P.O.Box 11100, FI-00076 AALTO, Finland}
\author{Patrick Rinke}
\affiliation{Department of Applied Physics, Aalto University, P.O.Box 11100, FI-00076 AALTO, Finland}
\maketitle

\section{Detail of C3 and C4 in the data screening scheme}
In C3, we excluded compounds that contain:
\begin{itemize}
  \item radioactive elements,
  \item toxic elements (Cd, Hg, Tl),
  \item noble metals as well as Re, Au and Te due to their very low abundance in earth's crust.
\end{itemize}
We also discarded all compounds containing H due to their high possibility of hydrolysis. In this work we only consider Ce to represent the whole rare-earth (lanthanides, scandium and yttrium) family as it is the most abundant member.

With the help of the CRC handbook for physics and chemistry \cite{crc} and other resources, we generated the data in C4 by manually discarding all compounds that are unstable in water, i.e., water soluble, reacting with water, or decomposing in water. However, the CRC handbook did not have solubility data on all our selected structures, leaving a huge set of data ($\sim$ 4000) mainly from the ternary class of compounds. We therefore need a more reliable reference to classify such compounds. In addition, compounds that have a very low melting point, are explosive, or easy to release poisonous gases were not included.

\section{Density-functional-theory calculations of halide perovskites}
DFT calculations were performed with the all-electron numeric-atom-centered orbital code \textsc{fhi-aims} \cite{blum09, havu, Levchenko/etal:2015}. For all calculations, we used \textit{tight} basis set and stress-tensor \cite{Knuth} implemented in \textsc{fhi-aims}. Scalar relativistic effect by means of the zero-order regular approximations (ZORA) \cite{vanlenthe} were included. The computational details were slightly different for the two perovskite classes considered in this work. We used the tetragonal P$_4$/mbm phase to construct $2\times2\times2$ supercell models for $\text{CsBX}_3^{}$, and relaxed the structures with a $\Gamma$-centered $4\times4\times4$ $k$-point mesh. While for $\text{MABX}_3^{}$, we used the tetragonal (quasi) I$_4$/mcm phase and optimized $\sqrt{2}\times\sqrt{2}\times2$ supercells with a $6\times4\times4$ $k$-point mesh. From the optimized tetragonal structures (with lattice constants $a,b=a,c$), we chose the $(001)$ planes to construct our virtual surface models for the lattice-mismatch calculations. The corresponding lattice constants are listed in Table~\ref{tab:lat}.

\begin{table}[!ht]
\caption{PBE-optimized lattice constants (in $\text{\AA}$) of $(001)$ surfaces of the investigated perovskites: single cells ($a_0^{}$), $2\times2$ surface cells ($a_{2\times2}^{}=2a_0^{}$) and $2\sqrt{2}\times2\sqrt{2}$ surface cells ($a_{2\sqrt[]{2}\times2\sqrt[]{2}}^{}=2\sqrt[]{2}a_0^{}$).}\label{tab:lat}
\vspace{-0.5em}
\begin{tabular}{
l@{\hspace{1.2em}}c@{\hspace{1.2em}}c@{\hspace{1.2em}}c@{\hspace{4em}}
l@{\hspace{1.2em}}c@{\hspace{1.2em}}c@{\hspace{1.2em}}c} \hline\hline
Cs- & $a_0^{}$ & $a_{2\times2}^{}$ & $a_{2\sqrt[]{2}\times2\sqrt[]{2}}^{}$ &
MA- & $a_0^{}$ & $a_{2\times2}^{}$ & $a_{2\sqrt[]{2}\times2\sqrt[]{2}}^{}$ \\ \hline
-SnCl$_3$ & 5.60 & 11.21 & 15.85 & -SnCl$_3$ & 5.76 & 11.52 & 16.29 \\
-PbCl$_3$ & 5.62 & 11.24 & 15.90 & -PbCl$_3$ & 5.82 & 11.64 & 16.46 \\
-SnBr$_3$ & 5.83 & 11.66 & 16.49 & -SnBr$_3$ & 5.99 & 11.98 & 16.95 \\
-PbBr$_3$ & 5.87 & 11.75 & 16.61 & -PbBr$_3$ & 6.01 & 12.01 & 17.00 \\
-SnI$_3$  & 6.18 & 12.37 & 17.49 & -SnI$_3$  & 6.30 & 12.61 & 17.83 \\
-PbI$_3$  & 6.23 & 12.46 & 17.63 & -PbI$_3$  & 6.36 & 12.72 & 17.99 \\ \hline \hline
\end{tabular}
\end{table}

\section{Suggested coating patterns for coatings of different lattice systems}

We only considered rectangular surfaces of coating materials therewith excluding the triclinic phases. Table~\ref{tab:lattsys} lists the lattice constants considered in the coating design (i.e., lattice-mismatch calculations). For hexagonal lattices (including trigonal lattices represented in hexagonal), we considered both $(10\bar{1}0)$ and $(11\bar{2}0)$ planes.

\begin{table}[!ht]
\caption{Lattice constants of cubic (\textsc{c}), tetragonal (\textsc{t}), orthorhombic (\textsc{o}), hexagonal (\textsc{h}), and monoclinic (\textsc{m}) coating materials considered in lattice-mismatch calculations.}\label{tab:lattsys}
\vspace{-0.5em}
\begin{tabular}{cl@{\hspace{1.em}}l} \hline\hline
& Lattice constants & Considered lattice constants 

\\ \hline
\textsc{c} & $a_{\text{c}}^{}$ & $a_{\text{c}}^{}$ 
\\
\textsc{t} & $a_{\text{c}}^{}=b_{\text{c}}^{},c_{\text{c}}^{}$ & $a_{\text{c}}^{}$ \\
\textsc{o} & $a_{\text{c}}^{},b_{\text{c}}^{},c_{\text{c}}^{}$ & ($a_{\text{c}}^{},b_{\text{c}}^{}$), ($a_{\text{c}}^{},c_{\text{c}}^{}$) or ($b_{\text{c}}^{},c_{\text{c}}^{}$) 
\\
\textsc{h} & $a_{\text{c}}^{}=b_{\text{c}}^{},c_{\text{c}}^{}$ & ($a_{\text{c}}^{},c_{\text{c}}^{}$),   ($\sqrt[]{3}{\text{a}}_{\text{c}}^{},c_{\text{c}}^{}$)  ($a_{\text{c}}^{},\sqrt{3}a_{\text{c}}^{}$) or ($ \sqrt[]{3}{\text{a}}_{\text{c}}^{}, \sqrt{3}{\text{c}}_{\text{c}}^{}$) 
\\
\textsc{m} & $a_{\text{c}}^{},b_{\text{c}}^{},c_{\text{c}}^{},\beta\ne90\text{\textdegree}$ & ($a_{\text{c}}^{},b_{\text{c}}^{}$) or ($b_{\text{c}}^{},c_{\text{c}}^{}$) 
\\ \hline\hline
\end{tabular}
\end{table}

\section{Coating parameters for \texorpdfstring{$\text{MAPbI}_3^{}$}{}}
We have listed the most suitable coating materials for $\text{MAPI}_3^{}$ with their lattice constants, lattice mismatch and their appropriate perovskite substrates in Tab.~\ref{tab:mapi3}. Here, we only listed the candidates with lattice mismatch $< 2$ \%. Binary and ternary coating materials with square planes are listed in (Tab.~\ref{tab:mapi3}a and (Tab.~\ref{tab:mapi3}b). We have also listed the lattice mismatch (along 2--axis) at the coating-$\text{MAPI}_3^{}$ interface for binary and ternary coatings with ``non-square`` planes (Tab.~\ref{tab:mapi3}c and d). 
\begin{table}[!htp]
\caption{Most suitable coating for $\text{MAPbI}_3^{}$ with $\vert\gamma\vert<5\%$: (a) \textsc{c} and \textsc{t} binary coatings, (b) \textsc{c} and \textsc{t} ternary coatings, and (c) \textsc{h}, \textsc{o} and \textsc{m} coatings. Shown are the lattice constants $a_{\text{c}}^{}$ (in $\text{\AA}$) of each coating considered in the $\gamma$ calculations, the lattice mismatch $\gamma$ (in \%) and the integer $m$ defined in Eqs.~(1) and (2) of the main text, and the size ($2\times2$ and $2\sqrt[]{2}\times2\sqrt[]{2}$, denoted by I and II, respectively) of perovskite (P) substrate.}\label{tab:result}
\begin{tabular}{l@{\hspace{1.3em}}c@{\hspace{1.3em}}c@{\hspace{1.3em}}c@{\hspace{1.3em}}c@{\hspace{1.3em}} c@{\hspace{1.3em}} c@{\hspace{3em}}
l@{\hspace{1.3em}}c@{\hspace{1.3em}}c@{\hspace{1.3em}} c@{\hspace{1.3em}}  c@{\hspace{1.3em}}c} 
\multicolumn{11}{c}{(a) \textsc{c}/\textsc{t} binary coatings / $\text{MAPbI}_3^{}$} \\
\hhline{=============}
& Structure & {$a_{\text{c}}^{}$}  & \multicolumn{1}{c}{$\gamma$} & $m$ & P &&
& Structure   & {$a_{\text{c}}^{}$} & \multicolumn{1}{c}{$\gamma$} & $m$ & P 
\\
\hline
MoF$_3$ & Pm\=3m   & $4.17$ & $-1.58$ & 4 & I &
 &NiO   & Fm\=3m & $4.23$ &  $-0.33
$ & 3 & I 
\\
SiO$_2$ & I\=43m     & $8.98$ & $-0.144$ & 1 & I &
& ZnO    & F\=43m     & $4.50
$ & $-0.01$ & 3 & II 
\\
GaN &  F\=43m   & $4.53$ & $0.65$ & 3 & II &
&BN & F\=43m        & $3.62$ & $0.73
$ & 4 & II 
\\
 ZrO$_2$ & P4$_2$/nmc & $3.59$ & $1.30
$ & 3 & II & &
 Si$_3$N$_4$    & I$_{43}$d     & $6.49
$ & $1.93$ & 3 & II
\\
ZnO & Fm\=3m     & $4.21$ &  $-0.75$ & 4 & II &
& BeO    & Fm\={3}m    & $3.65$ &  $1.39$ & 3 & I
\\
SiO$_2$     & F\=3m     & $4.58$ &  $1.88$ & 3 & I &
& BN & P4$_2$/mmc & $9.13$ &  $1.46$ & 1 & I
\\
\hhline{=============}
\vspace{-0.5em} \\
\multicolumn{11}{c}{(b) \textsc{c}/\textsc{t} ternary coatings / $\text{MAPbI}_3^{}$} \\
\hhline{=============}
& Structure & {$a_{\text{c}}^{}$}  & \multicolumn{1}{c}{$\gamma$} & $m$ & P &&
& Structure   & {$a_{\text{c}}^{}$} & \multicolumn{1}{c}{$\gamma$} & $m$ & P \\
\hline
 PbZrO$_3$ &  Pm\=3m  & $4.21$ & $-0.75
$ & 4 & II & 
 & CaZrO$_3$ &  Pm\=3m   &  $4.16$ &  $-1.89$ & 4 & II
 \\
 ZrW$_2$O$_8$ & P2$_1$3     &  $3.54
$ &  $-1.66$ & 1 & I &
 & SrZrO$_3$ & P2$_1$3      &  $3.55
$ & $-1.41$ & 5 & I 
\\
CaSiO$_3$ & Pm\=3m      &  $3.61$ & $0.27$ & 4 & I &
& BaAl$_2$S$_4$ & Pa\=3    &  $3.65$ & $1.40$ & 3 & I
\\
SrZrO$_3$ &P4/mbm &$3.17$ &$-0.42$ & $4$ &II &
& PbTiO$_3$ &I4/m &$12.71$ &$-0.10$ & $1$ &II \\
\hhline{=============}
\vspace{-0.5em} \\
\multicolumn{11}{c}{(c) \textsc{h}/\textsc{o}/\textsc{m}/\textsc{r} binary coatings / $\text{MAPbI}_3^{}$} \\
\hhline{=============}
& Structure & {$a_{\text{c}}^{}$}  & \multicolumn{1}{c}{$\gamma$} & $m$ & P &&
& Structure   & {$a_{\text{c}}^{}$} & \multicolumn{1}{c}{$\gamma$} & $m$ & P \\
\hline \\
BN  & R3m &  $2.51$ &  $-1.27$ & 5 & I && BN &P6$_3$mc &$2.56$ &$0.42$ &5 &I
 \\
      &            & $12.16$ & $^{\ast}-4.59$ & 1 & I && &&$4.23$ & $-0.36$ & 3 & I
 \\
  SiO$_2$    & P6$_5$22  &  $7.36$ &  $0.20$ & $\sqrt{3}$ & I &&  SiO$_2$  & Cmca   &  $9.09$ & $1.01$ & 2 & I 
\\
            &        &  $7.06$ &  $^{\ast}-4.07$ & $\sqrt{3}$ & I & 
            & &                  & $9.36$ & $3.90$ & 2 & II
\\
SiC & P3m1   &  $3.09$ &  $-2.79$ & 4 & I 
 & &&&&&&\\ && $12.66$ &  $-0.52$ & 1 & I
\\ 
\hhline{=============}

\vspace{-0.5em} \\
\multicolumn{11}{c}{(d) \textsc{h}/\textsc{o}/\textsc{m}/\textsc{r}  ternary coatings / $\text{MAPbI}_3^{}$} \\
\hhline{=============}
& Structure & {$a_{\text{c}}^{}$}  & \multicolumn{1}{c}{$\gamma$} & $m$ & P &&
& Structure   & {$a_{\text{c}}^{}$} & \multicolumn{1}{c}{$\gamma$} & $m$ & P \\
\hline \\
Al$_2$SiO$_5$ & Cmcm  &  $3.56$ &  $-0.96$ & 5 & II & & BaSiF$_6$ & R\=3m  &  $7.33$ &  $-0.23$ & $\sqrt{3}$ & I 
\\
         &       & $9.29$ & $1.96$ & 2 & II & &            &       & $7.12$ & $^{\ast} -3.20$ & $\sqrt{3}$ & I
\\
MgSiO$_3$  & Pbcn  &  $8.86$ &  $-1.59$ & 2 & II && MgAl$_2$O$_4$ & Cmcm &2.81 & $2.99$ & 6 &II 
\\   &  & $9.36$ & $3.90$ & 2 & II & &&& $9.27$
& $^{\ast}4.94$&2 &II
\\
MgSiO$_3$ &Pbca &$8.86$ &$-0.61$ &2 &II  &&BaGeF$_6$ & R\=3m &$7.48$ &$1.82$ & $\sqrt{3}$ &I \\
 & &$18.45$ &$2.46$ &$1$ & II & & & &$7.24
$ &$-1.45
$ &$\sqrt{3}$ &I
\\
\hhline{=============}
\multicolumn{6}{l}{$^{\ast}$: {\footnotesize $=\sqrt[]{3}\cdot3.62$}}
\end{tabular}\label{tab:mapi3}
\end{table}

\section{Coating materials with multiple phases}
\begin{figure}[!htp]
\centering
\includegraphics[width=17.5cm]{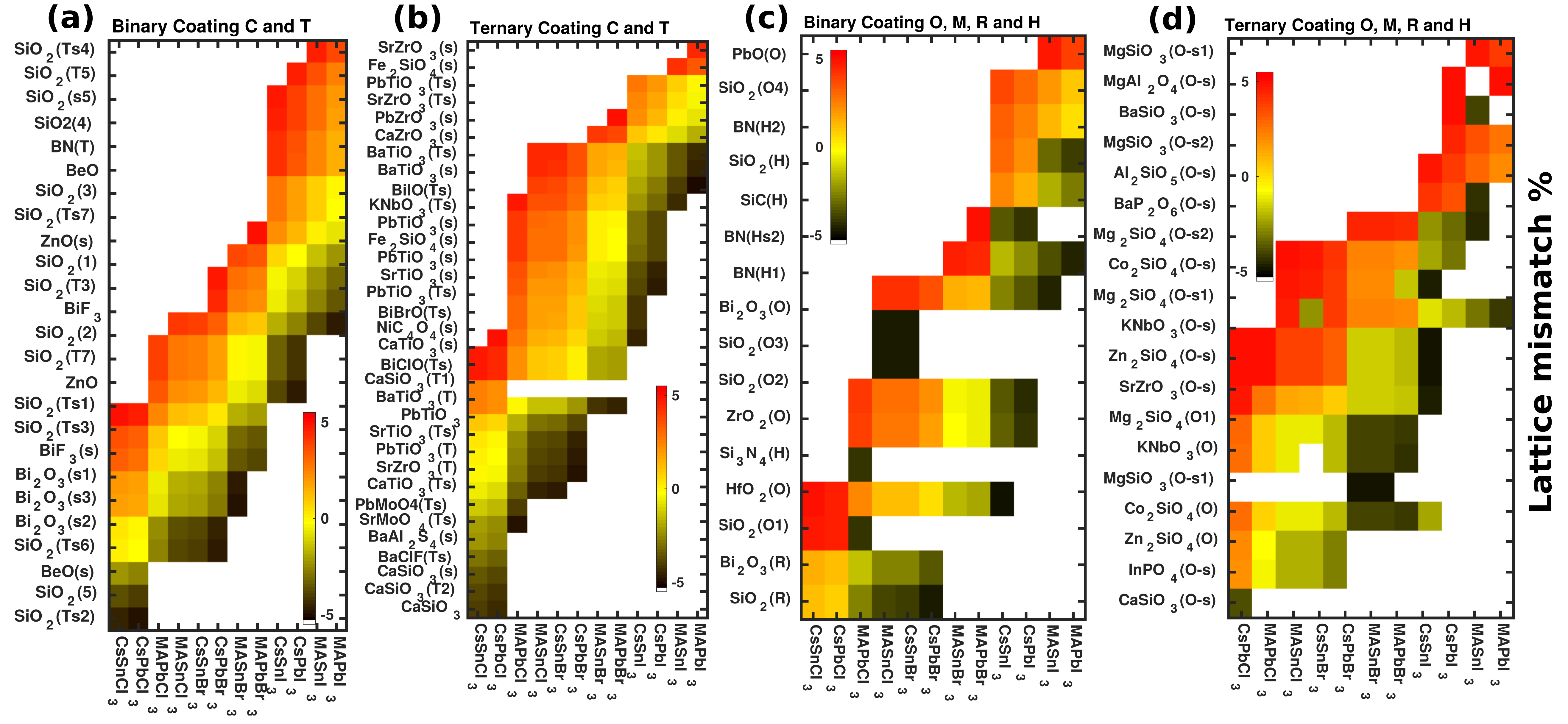}
\caption{Calculated lattice mismatch ($\gamma$) between considered perovskites (horizontal axes) and all coating materials (vertical axes) outcome from the screening scheme:  \textsc{c},  \textsc{t},   \textsc{o}, \textsc{m}, \textsc{h} and \textsc{r} represent, cubic, tetragonal, orthorhombic, monoclinic, hexagonal and rhombohedral crystal structures, respectively. `s` denotes the $2\sqrt{2}\times2\sqrt{2}\times2\sqrt{2}$ perovskite substrates. (a) binary \textsc{c} (unlabeled) and \textsc{t} coatings, (b) ternary \textsc{c} and \textsc{t} coatings, (c) and (d) ``non-square'' (i.e., \textsc{o}, \textsc{m}, \textsc{h} and \textsc{r}) for binary and ternary coatings. Numbers 1, 2 ... denote materials of the same crystal structures having different phases}\label{fig:coat} 
\end{figure}

Suitable coating materials for PSCs with  multiple phases are shown in \ref{fig:coat}. Here, we show all the remaining coating candidates that were not shown in the main text. We also show the possible coating candidates for both $2\times2$ and $2\sqrt{2}\times2\sqrt{2}$ perovskite substrates used in this work. Here, we use ``s`` to denote coating materials for the $2\sqrt{2}\times2\sqrt{2}$ surfaces. Cubic phases are not labelled, ``$\textsc{t}$``, ``$\textsc{o}$``, ``$\textsc{r}$`` and ``$\textsc{h}$`` denote tetragonal, monoclinic, rhombohedral and hexagonal structures respectively. We used the numbers, $1, 2, 3, ...$ to differentiate between coating materials of the same structures but different phases. 

\newpage
\bibliographystyle{apsrev}
\bibliography{sample}